# Intercomparison of Monte Carlo calculated dose enhancement ratios for gold nanoparticles irradiated by X-rays: assessing the uncertainty and correct methodology for extended beams

H. Rabus[1,13], W.B. Li[2,13], C. Villagrasa[3,13], J. Schuemann[4,13], P.A. Hepperle[1,5], L. de la Fuente Rosales[1], M. Beuve[6,13], S. Di Maria[7,13], A.P. Klapproth[2,8], C.Y. Li[9,10], F. Poignant[6,11], B. Rudek[1,4,12], H. Nettelbeck[1,13]

[1] *Physikalisch-Technische Bundesanstalt, Braunschweig and Berlin, Germany*
[2] *Institute of Radiation Medicine, Helmholtz Zentrum München - German Research Center for Environmental Health, Neuherberg, Germany*
[3] *Institut de Radioprotection et de Sûreté Nucléaire, Fontenay-Aux-Roses, France*
[4] *Massachusetts General Hospital & Harvard Medical School, Department of Radiation Oncology, Boston, MA, USA*
[5] *Leibniz Universität Hannover, Hannover, Germany*
[6] *Institut de Physique des 2 Infinis, Université Claude Bernard Lyon 1, Villeurbanne, France*
[7] *Centro de Ciências e Tecnologias Nucleares, Instituto Superior Técnico, Universidade de Lisboa, Bobadela LRS, Portugal*
[8] *TranslaTUM, Klinikum rechts der Isar, Technische Universität München, Munich, Germany*
[9] *Department of Engineering Physics, Tsinghua University, Beijing, China*
[10] *Nuctech Company Limited, Beijing, China*
[11] *NASA Langley Research Center, Hampton, VA, USA*
[12] *Perlmutter Cancer Center, NYU Langone Health, New York City, NY, USA*
[13] *European Radiation Dosimetry Group (EURADOS) e.V, Neuherberg, Germany*

ABSTRACT

Results of a Monte Carlo code intercomparison exercise for simulations of the dose enhancement from a gold nanoparticle (GNP) irradiated by X-rays have been recently reported. To highlight potential differences between codes, the dose enhancement ratios (DERs) were shown for the narrow-beam geometry used in the simulations, which leads to values significantly higher than unity over distances in the order of several tens of micrometers from the GNP surface. As it has come to our attention that the figures in our paper have given rise to misinterpretation as showing 'the' DERs of GNPs under diagnostic X-ray irradiation, this article presents estimates of the DERs that would have been obtained with realistic radiation field extensions and presence of secondary particle equilibrium (SPE). These DER values are much smaller than those for a narrow-beam irradiation shown in our paper, and significant dose enhancement is only found within a few hundred nanometers around the GNP. The approach used to obtain these estimates required the development of a methodology to identify and, where possible, correct results from simulations whose implementation deviated from the initial exercise definition. Based on this methodology, literature on Monte Carlo simulated DERs has been critically assessed.

## 1. Introduction

GNPs are considered as potential radiosensitizers in the field of radiation therapy [1], [2], [3], [4], [5], [6], as they have been shown to enhance the biological radiation effectiveness in vitro and in vivo [1], [6], [7], [8], [9]. This effect is often explained by a local enhancement of absorbed dose around GNPs that results from the higher absorption of radiation by the high-Z material gold as compared to biological matter or water. Due to the stronger interaction with the incoming photons and ensuing Auger cascades, a larger number of secondary electrons will be emitted from the GNP and deposit additional energy in its vicinity as compared to the case when the GNP volume is also filled with water. The ratio of the absorbed dose within water volumes surrounding the GNP for the two cases of GNP present or absent is the dose enhancement ratio (DER). As absorbed dose is a point quantity, so is the DER, which shows large variation in the close vicinity of the GNP [10], [11], [12], [13],

[14]. The additional energy deposition also results in an enhancement of the average dose within larger volumes, e.g. a cell or a tumor volume when they are loaded with GNPs. It is therefore also common practice to consider the DER of the average absorbed dose induced by GNPs in such larger volumes [15], [16], [17], [18], [19], [20], [21], [22].

The dose enhancement from GNPs is limited to microscopic dimensions and, therefore, requires determination by theoretical methods or by Monte Carlo (MC) simulations. The small size of GNPs of typically a few tens of nanometers imposes a major challenge on such MC simulations.

As shown in a recent review by Moradi et al. [22], more than 120 articles on nanoparticle "radiosensitization" by Monte Carlo methods have been published in the last 15 years (including their paper and the recent review by Vlastou at al. [21]). Depending on whether dose enhancement is assessed on the nanometric scale around a single nanoparticle or averaged over cellular or larger dimensions (i.e. several 10 µm to several mm), DER results reported in the literature range from values close to unity to several thousand [19], [21], [22].

As the discrepancies in literature regarding DERs on the sub-cellular scale could in principle be due to different cross section data sets employed by different codes, a code intercomparison exercise was conducted as a joint activity of Working Groups 7 "Internal Dosimetry" and 6 "Computational Dosimetry" of the European Radiation Dosimetry Group [23], [24], [25]. The exercise was an intercomparison of Monte Carlo simulations for the case of dose enhancement in water from two sizes (50 nm and 100 nm diameter) of single spherical gold nanoparticles (GNP) under X-ray irradiation [26], [27].

The exercise aimed at studying the dependence of the results on both the MC codes used and the choice of adjustable parameters (cut-off energies, cross section models, etc.; for details see Supplementary Tables 1 and 2). For this purpose, a photon beam with a cross sectional area comparable to that of the GNP was used in the simulations (see Fig. 1). In this way, differences between codes such as in the interaction cross sections and implementation of electron transport in gold are emphasized. Furthermore, the range of photon energies was chosen such as to have maximum differences between photon interaction cross sections for gold and water.

Participants in the exercise were requested to simulate the irradiation of a single spherical gold nanoparticle surrounded by water with a parallel photon beam of circular cross section as illustrated in Fig. 1. The photon source was placed at a distance of 100 μm from the GNP and had a diameter of 60 nm or 110 nm for a GNP diameter of 50 nm or 100 nm, respectively. Two X-ray spectra were considered that corresponded to tube voltages of 50 kVp and 100 kVp (see Supplementary Table 3).

For all combinations of photon energy spectrum and GNP size, simulations were performed (a) with the GNP in place and surrounded by water and (b) with the respective volume only filled with liquid water. Energy deposition was scored in spherical water shells around the GNP volume. Up to 1 μm from the GNP surface, the spherical shells had a thickness of 10 nm, and between 1 μm and 50 μm the shell thickness was 1 μm. DERs were determined as the ratio of the energies deposited in each spherical shell in the simulations with GNP and for water only. In addition, participants were also requested to report the energy spectra of electrons emitted from the GNP for all combinations of GNP size and photon spectrum.

The DERs presented in [26], [27] were significantly higher than unity over distances in the order tens of micrometers from the GNP surface. For instance, in Fig. 6(d) in [26], [27] all reported DERs for the 100 nm GNP and 100 kVp X-ray spectrum are higher than 1.5 up to 20 μm distance. This could be interpreted as the presence of a single 100 nm GNP during irradiation leading to a more than 50% increase of absorbed dose in a cell. However, this is unrealistic as (assuming a cell diameter of 10 μm) the mass fraction of gold (MFG) would be about $2 \times 10^{-5}$, and thus, with up to 100 times higher mass energy absorption of gold compared to water [28], the dose enhancement cannot exceed the order of 0.2%.

It is well known that results for DERs are highly dependent on the simulation setup [29], [30], [31] and it was clearly stated in [26] that the shown DERs are valid only for the particular exposure scenario that was used for the simulations. As it has come to our attention that the results reported in [26], [27] are prone to be misinterpreted at representing 'the' DERs of GNPs under diagnostic X-ray irradiation, this paper shows how much different the DERs would be for a GNP irradiated with an extended photon field.

## 2. Materials and Methods

### 2.1. Physical background

If simulations are carried out using a photon beam of microscopic cross section, then the energy imparted in the scoring volumes is predominantly deposited by electron tracks produced within the region traversed by the primary photon beam. In the exercise of [26], this region was a water cylinder of about 100 μm height (diameter of the scoring region) and a diameter equal to the GNP diameter plus 10 nm. Thus, the MFG was in the order of 1% to 2% and this high MFG was the reason for the large DERs and extended spatial range, where DERs were significantly larger than unity.

In an extended photon field, electrons generated by photon interactions outside the volume, which is defined by the beam direction and GNP cross section, also deposit energy in the vicinity of the GNP. In case of secondary charged particle equilibrium (CPE), that is if the energy transported out of a volume by leaving electrons equals the energy introduced by electrons entering the volume, the absorbed dose is equal to the energy released in photon interactions in that volume.

Over microscopic dimensions where the attenuation of the photon beam can be neglected, the absorbed dose then has a constant value. In the presence of the GNP, the conditions of CPE are not met for the volume of the GNP, as the gold atoms are greater absorbers than water molecules, and thus, more energy is transported by electrons out of the GNP volume than into it.

For an extended photon beam, photons scattered outside the confined volume (defined by beam direction and GNP cross section) can also reach the GNP and have interactions with it or within its vicinity. The electrons released by interactions of these scattered photons in the GNP or its surroundings increase the energy deposited in the regions of interest.

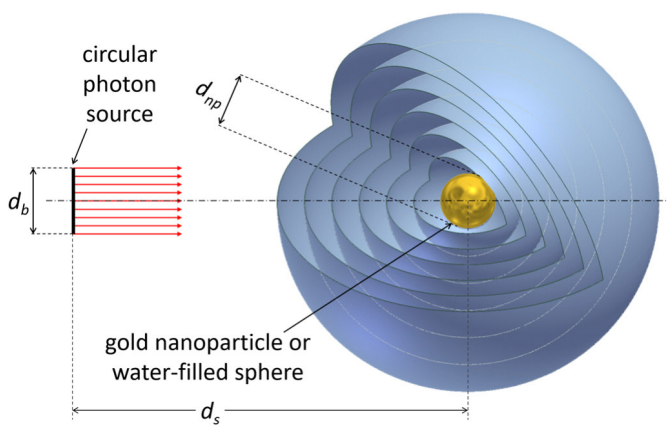

**Fig. 1:** Schematic illustration of the geometry setup for the simulations. A homogeneous circular X-ray source of diameter $d_b$ emits photons along its rotational symmetry axis (dot-dashed line). Energy deposition is scored in concentric spherical shells around a sphere of diameter $d_{np}$ located at a distance $d_s$ = 100 μm from the photon source. This sphere is filled with either gold or water; the rest of the geometry is filled with water. The first 100 spherical shells have a radial thickness of 10 nm; the remaining 49 spherical shells have a radial thickness of 1 μm. The diameter of the photon source is $d_b = d_{np} + 10$ nm; the diameter of the central sphere (gold nanoparticle) is either 50 nm or 100 nm.

Hence, in simulations with confined beams, the lack of CPE together with the absence of interactions of scattered photons in the water volume around the GNP leads to an increase in the DER, while neglecting the interactions of scattered photons with the GNP leads to a slight reduction in the DER.

A procedure to use the results of such simulations to derive the DER for an extended photon field was described in detail in [32]. It is based on two assumptions that are briefly reviewed in the next two subsections: one is related to the correction for the lack of charge particle equilibrium (CPE) and one to the lack of secondary photon equilibrium (SPE).

### 2.2. Correction for lack of charged particle equilibrium (CPE)

The first assumption is that the lack of energy deposition in the water shells around the GNP from electrons originating from outside the confined beam geometry is the same with and without the GNP present. This assumption means that the difference of energy per mass obtained from the simulations with and without the GNP can be used as value for the extra contribution coming from the GNP. This contribution then adds to the dose to water, $D_w$, in the absence of the GNP, which on the conditions of CPE is equal to the collision kerma [33]

$$D_w = \int E \times \frac{\mu_{en,w}(E)}{\rho} \times \Phi(E) e^{-\mu_w(E)d_g} dE \quad (1)$$

where $E$ is the photon energy, $\mu_{en,w}(E)/\rho$ is the mass energy absorption coefficient of water, $\Phi(E)$ is the spectral fluence (particles per area and energy interval) of photons in the region of interest (i.e. around the GNP position), $\mu_w(E)$ is the total linear attenuation coefficient of water and $d_g$ is the distance of the GNP volume center from the photon source. As the attenuation of the photon beam is negligible over the microscopic dimensions considered, the absorbed dose to water is constant in the absence of the GNP.

In the presence of the GNP, there is additional energy deposition from electrons emerging from the GNP, while the energy deposition from electrons produced in the same volume in water when the GNP is absent is missing. The simulation results of deposited energy in the spherical shells around the GNP are basically the sum of aforementioned contributions and the energy depositions from electrons released by photon interaction in the rest of the volume traversed by the beam (except the sphere covered by the GNP or filled with water). Therefore, the difference of the simulation results for the cases with and without the GNP can be used as approximation for the extra dose contribution due to the GNP.

With this approximation, the CPE-corrected dose $D_g^{(p)}(r)$ in a spherical shell between radial distances $r$ and $r+\Delta r$ from the GNP surface is approximately given by

$$D_g^{(p)}(r) = D_w^{(p)} + \frac{\bar{\varepsilon}_g(r) - \bar{\varepsilon}_w(r)}{m_w(r)} \quad (2)$$

In eq. (2), $D_w^{(p)}$ is the dose to water under CPE conditions calculated with eq. (1) using the primary photon spectral fluence $\Phi^{(p)}$ with the normalization condition

$$\int \Phi^{(p)}(E) dE = \frac{1}{r_b^2 \pi} \quad (3)$$

where $r_b$ is the radius of the circular photon source used in the simulations. $\bar{\varepsilon}_g(r)$ and $\bar{\varepsilon}_w(r)$ in eq. (2) are the average imparted energies per primary photon obtained in the simulations with the GNP and without the GNP, respectively, and $m_w(r)$ is the mass of water in the spherical shell, which is approximately given by

$$m_w(r) = \rho_w 4\pi (r_g + r)^2 \Delta r \quad (4)$$

where $\rho_w$ is the mass density of water and $r_g$ is the GNP radius.

### 2.3. Correction for lack of scattered photon equilibrium (SPE)

The second assumption is that the excess dose $D_g^{(s)}$ due to photons produced (in an extended photon beam) by scattering interactions outside the confined beam geometry that interact with the GNP can be estimated by multiplying the difference between simulations with and without the GNP using a scaling factor. This scaling factor is the ratio of the collision kermas in gold for the energy spectra of scattered photons and for the primary photon spectrum, $D_{Au}^{(s)}$ and $D_{Au}^{(p)}$, respectively, i.e.

$$D_g^{(s)}(r) = D_w^{(s)} + \frac{\bar{\varepsilon}_g(r) - \bar{\varepsilon}_w(r)}{m_w(r)} \frac{D_{Au}^{(s)}}{D_{Au}^{(p)}} \quad (5)$$

where $D_w^{(s)}$ is the dose to water calculated with eq. (1) using the spectral fluence of scattered photons, $\Phi^{(s)}$, and the collision kerma in gold is calculated according to

$$D_{Au} = \int E \times \frac{\mu_{en,Au}(E)}{\rho} \times \Phi(E) dE \quad (6)$$

where $\mu_{en,Au}(E)/\rho$ is the mass energy absorption coefficient of gold. The approximation is that the dose deposited in water around the GNP scales with the collision kerma of photon interactions inside the GNP. As the electrons released by photon interactions in the GNP lose part of their energy in the GNP, there may be a dependence on the size of the GNP and the photon energy spectrum.

The spectral fluence of scattered photons that reach the GNP can be determined analytically only for the case that exactly one coherent or incoherent scattering event has occurred. The fluence of all generations of scattered photons that reach the GNP is estimated from the ratio of the sum of spectral fluences of all generations of scattered photons to the spectral fluence of the first generation that would be obtained if photon scattering was isotropic. This ratio is multiplied by the spectral fluence of singly scattered photons with the correct angular dependence to approximate the fluence of all generations of photons scattered with the proper angular dependence [32].

The spectral fluence of singly scattered photons can be calculated by multiplying the following probabilities and integrating over the volume traversed by the extended primary photon beam:

(1) Probability that a primary photon reaches the interaction point without any preceding interactions and that the scattered photon reaches the GNP without further interaction. Both probabilities are given by the Lambert-Beer law using the total photon mass attenuation coefficient that is available from literature [34].

(2) Probability that a coherent or incoherent scattering event occurs per unit length, which is given by the respective mass attenuation coefficients [34].

(3) Probability that the scattering occurs within the solid angle covered by the GNP. This can be estimated from the differential Thomson and Klein-Nishina cross sections [32].

The photon energy spectrum obtained by calculating the fluence of the first generation of scattered photons and scaling to all generations of scattered photons is then used in eq. (1) to calculate the collision kerma due to photon interactions in the volume of the GNP filled with either water or gold by using the respective mass-energy absorption coefficients. Under CPE conditions, the collision kerma obtained for the volume filled with water is equal to the additional dose to water due to the 'additional' irradiation with the scattered photons.

The collision kerma obtained for the volume filled with gold is the energy per mass transferred to electrons by photon interactions in gold. The ratio of this collision kerma value (due to the spectrum of scattered photons) and the collision kerma for gold calculated with the primary photon energy spectrum is the factor used to multiply the contribution of the absorbed dose around the GNP (due to photon interactions within the GNP as obtained from the simulations) to estimate the excess dose contribution from electrons released in the GNP by scattered photons [32].

*2.4. Consistency checks for the simulation results*

For the narrow-beam geometry used in the exercise, the ratio of simulation results with and without GNP present is almost insensitive to normalization as long as it is done consistently in the two simulations for each combination of GNP size and spectrum. On the contrary, the application of the correction for CPE and SPE requires that the simulated data are related to a known primary photon fluence. To ensure that this is the case, the ensemble of reported data (i. e. energy spectra of electrons emitted from the GNP and energy deposition in the spherical shells around it for all combinations of GNP size and photon spectrum) of each participant was subjected to plausibility and consistency checks.

The first check was a test for the overall normalization. This was based on the ratio between the total energy deposited in a sphere of radius $R$ in simulations without a GNP and the average energy transferred by photon interactions in water occurring in the part of the sphere traversed by the primary photon beam. For large values of $R$, the condition of secondary electron equilibrium along the direction of the primary photon beam is fulfilled, such that the ratio should converge to a value close to unity.

The second test was to check for consistency between emitted electron energy spectra and radial integral of deposited energy around the GNP.

The third plausibility check was to examine the consistency between the data for different GNP sizes and photon energy spectra. This required testing whether the ratio of the radial integral of the average additional energy deposited in the presence of a GNP, $\Delta E_{d,g}$, and the expected number of photon interactions in the GNP, $\bar{n}_g$, was compatible with the average energy transferred to electrons when a photon interacts in gold, $E_{tr}$. $\bar{n}_g$ was calculated according to

$$\bar{n}_g = \frac{4\pi}{3} r_g^3 \int \mu_{Au}(E) \times \Phi^{(p)}(E) e^{-\mu_w(E) d_g} \, dE \qquad (7)$$

and, thus, depends on the GNP size and photon energy spectrum.

As electrons released in photon interactions within the GNP lose part of their energy before leaving the GNP, the $\Delta E_{d,g}/\bar{n}_g$ must be smaller than $E_{tr}$. Furthermore, for the same photon spectrum, the ratio $\Delta E_{d,g}/\bar{n}_g / E_{tr}$ should be smaller for the 100 nm GNP than for the 50 nm GNP, since the average path travelled by electrons before leaving the GNP is smaller for the smaller GNP. For the same GNP size, $\Delta E_{d,g}/\bar{n}_g / E_{tr}$ for the 100 kVp spectrum should be smaller than that for the 50 kVp spectrum, since electrons produced by photo-absorption in the L, M and outer shells and Compton scattered electrons have higher energies. The 100 kVp photon spectrum also contains photon energies where K shell absorption is possible. The fraction of such photons is small and the photo-absorption coefficient around the K shell of gold is lower than in the photon energy range below 50 keV where the majority of photons in the spectrum lie [34].

For those sets of results where the consistency tests indicated specific normalization issues, the respective participants were requested to check and confirm whether their simulations were compromised by the respective problem. Examples include improper implementation of the simulation geometry, such as using a source where the radius was larger than the GNP radius by 10 nm rather than applying this difference for the diameters. If the participant confirmed that the simulations were biased as suggested by the outcomes of the consistency checks, the results were corrected accordingly.

A detailed account of this analysis is beyond the scope of this paper and will be published as a comprehensive EURADOS report on the lessons learned from the exercise. For the energy spectra of electrons emitted from the GNP, where the identified normalization issues affect the data shown in Figs. 7 and 8 of our previous paper [26], a corrigendum has been published [27].

For the determination of the best estimate for the DER in an extended photon field as well as for the associated uncertainty, we considered in this work only those datasets that passed (from the start or after revision) all three tests described in the next three sub-sections. The best estimate was taken as the mean of these data sets and the associated expanded uncertainty at 95% confidence level was taken as the geometrical mean of twice the sample standard deviation among the accepted results and an estimate for the uncertainty of the corrections (see Section 3.2).

## 3. Results

*3.1. Application of corrections to the simulated data*

The resulting corrections are illustrated in Fig. 2 for the case of a 100 nm GNP irradiated with the 50 kVp X-ray beam, i.e. the combination for which the largest DER values were obtained by all participants in the exercise [26]. In Fig. 2, we use the data of a participant whose results were generally around the median of the reported DERs in [26] for each combination of X-ray spectrum and GNP size.

In all three panels of Fig. 2, the dashed line indicates the absorbed dose to water in the absence of the GNP if CPE exists and if only the primary photon spectrum is considered. The solid line is the absorbed dose to water calculated with the estimated fluence of all generations of scattered photons, i.e. under SPE conditions. The results obtained from the simulations with and without the GNP are shown in Fig. 2(a) as symbols (see legend). The difference between these results is shown in Fig. 2(b) as downward triangles, which is approximately equal to the dose contribution in the surrounding water arising from electrons emitted by the GNP following photon interactions. The correction obtained after adding the estimated contribution from scattered photons is indicated by the circles.

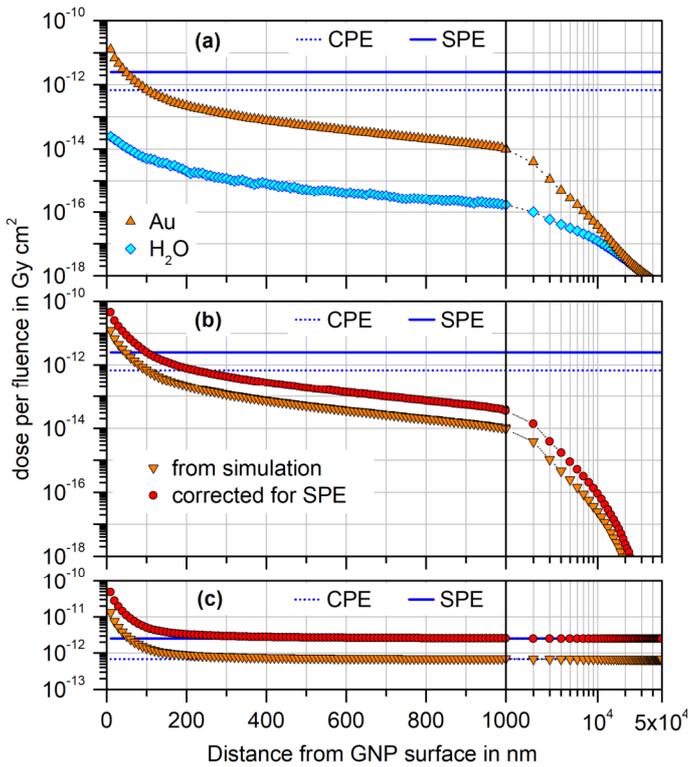

**Fig. 2:** (a) Simulation results of participant G1 for dose per fluence in 10 nm-thick spherical shells at different radial distances from a 100 nm gold nanoparticle (GNP) irradiated in water with a circular 50 kVp photon beam of diameter 110 nm (symbols). The *x*-axis shows the difference between the radius of the spherical shell's outer surface and the GNP radius and is linear up to 1000 nm and then logarithmic. The dashed blue lines on all panels indicate the value obtained without GNP for charged particle equilibrium (CPE); the solid blue lines mark the dose per fluence of primary particles for the case of full secondary particle equilibrium (SPE), i.e. including scattered photons. (b) Difference of the simulation results in (a) for the GNP present and absent (downward triangles) and corrected values taking into account scattered photons (circles). (c) Sums of the data shown in (b) and the constant values of dose to water per fluence for CPE and SPE.

Finally, Fig. 2(c) shows the results obtained by adding the estimated dose contributions due to the GNP and the respective estimated dose to water values in the absence of the GNP, when only the correction for CPE is applied (downward triangles) and after accounting also for the correction for SPE (circles).

The DER calculated from the simulation results presented in Fig. 2(a) for the confined beam is shown in Fig. 3 as upward triangles. The squares indicate the DERs obtained after applying the correction for CPE according to eq. (2), and the circles (which are mostly overlapping the squares) are the results from applying the additional correction for scattered photons. From this graph, it is evident that in an extended photon field significant dose enhancement is only found within the first few hundred nanometers around the GNP, and that scattered photons interacting with the GNP do not significantly change the resulting DER.

Fig. 4 shows a comparison between the DERs for a narrow beam and for an extended beam geometry in the range up to 1000 nm from the GNP surface for all combinations of GNP size and X-ray spectrum. The lin-log plots allow a direct comparison with Fig. 4 in [26]: In contrast to the significant dose enhancement over micrometers for the confined beam, the DER drops rapidly towards unity over the first few hundred nanometers from the GNP surface.

In Fig. 4 all available data have been included, where data that failed to pass the plausibility tests are marked by open symbols. It can be seen that these data are at the extremes of the spread of results, which suggests that corrections of the unidentified normalization issues might further reduce the deviations between the results, both for the narrow beam and for the extended beam with CPE and SPE conditions.

In [26], the uncertainties for the DERs for the confined photon beam were estimated from a two-sided 95% coverage interval for the spread of results obtained by different participants (assuming a log-normal distribution) that was interpreted as the 95% confidence interval. The upper limit of this confidence interval was typically larger than the lower limit by about a factor of two. As in Fig. 4 the spread of results for the extended beam is much smaller than for the narrow beam geometry, a normal rather than a log-normal distribution may be assumed when estimating the uncertainty of the DER of an extended photon beam from the scatter of results.

However, in this case the corrections for CPE and SPE and the approximations that they imply constitute another source of uncertainty to be accounted for. Fig. 2 shows that the applied CPE and SPE corrections change the dose-per-fluence ratios by more than three orders of magnitude. Hence, the uncertainties associated with the corrections may be expected to be substantial and are discussed in the next Section.

### 3.2. Uncertainty of the CPE and SPE corrections

The first assumption, which is the basis of the correction for lack of CPE, neglects the so-called sink effect of the GNP [35], i.e. that electrons traversing the GNP will lose more energy than when traversing the same volume of water. This will lead to a reduction of the energy deposited in the water around the GNP. On the other hand, if the electron energy is high enough, it may also ionize core levels of gold and create an Auger cascade that enhances the energy deposition in the water around the GNP. However, taking into account the underestimation of one or more orders of magnitude of absorbed dose to water without the GNP, as seen in Fig. 2(a), the uncertainty introduced by neglecting these two effects is probably insignificant.

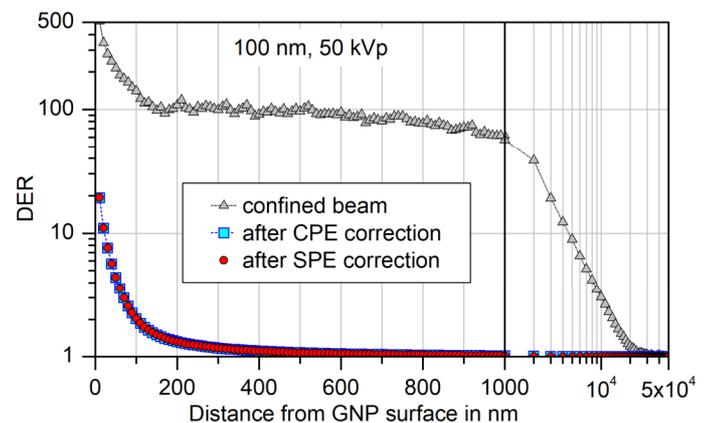

**Fig. 3:** Dose enhancement ratio (DER) for a 100 nm GNP and 50 kVp X-rays calculated from the data shown in Fig. 2(a) and Fig. 2(c), respectively. The symbols for the results after correcting for CPE (boxes) and after also correcting for SPE (circles) are overlapping.

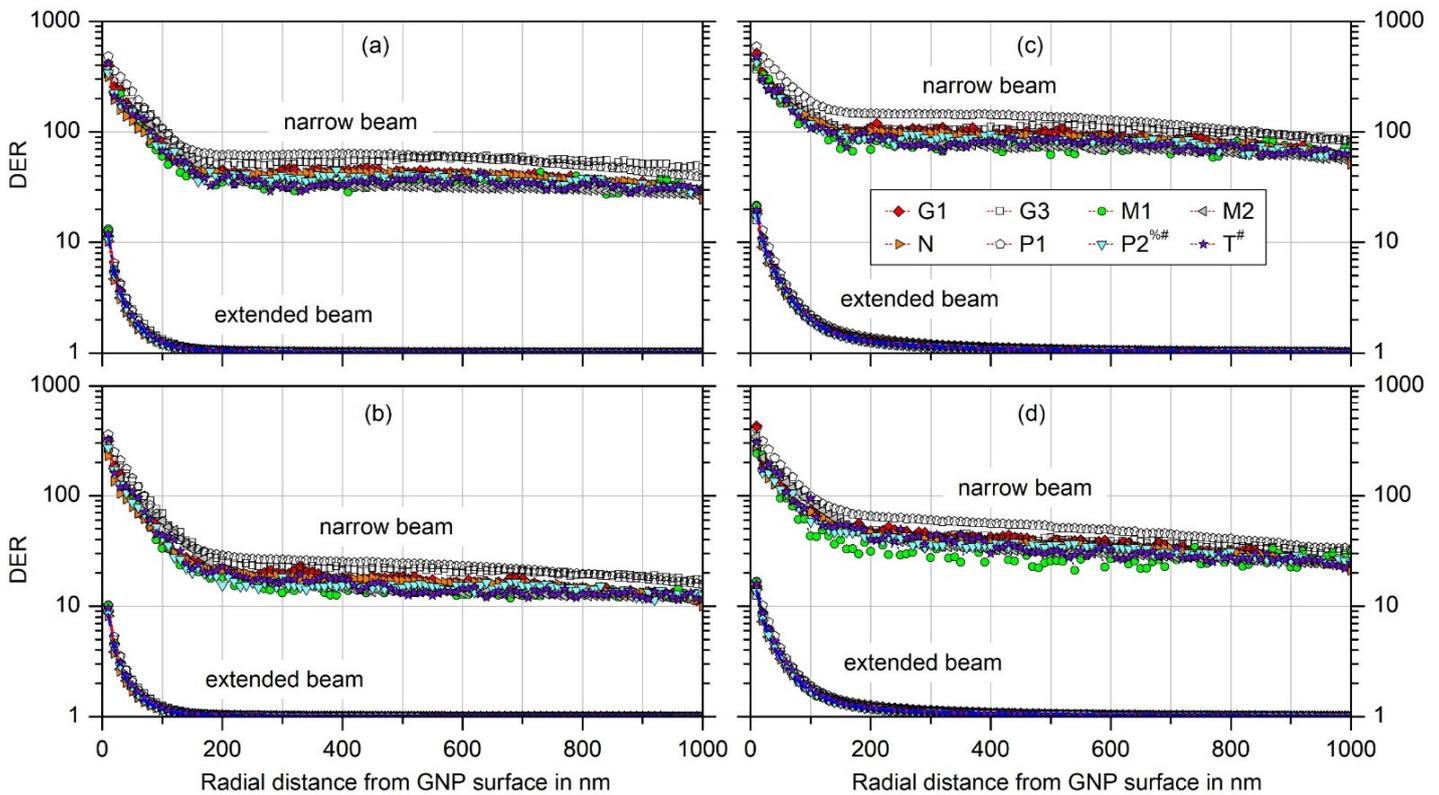

**Fig. 4**: Comparison of the dose enhancement ratios (DER) for the narrow beam used in the simulations in [26] and for an extended photon field under secondary particle equilibrium for (a) a 50 nm GNP and 50 kVp X-rays, (b) a 50 nm GNP and 100 kVp X-rays, (c) a 100 nm GNP and 50 kVp X-rays and (d) a 100 nm GNP and 100 kVp X-rays. The different symbols correspond to the different participants in the exercise. Open symbols indicate that the reported simulation results did not pass the plausibility checks and no explanation could be reported by the participant. The other data sets passed the plausibility tests, where in some cases identified normalization issues have been corrected for. A hashtag on the participant identifier indicates that a fluence correction was applied on the data (as the photon source geometry was at variance with the exercise definition). A percent sign indicates that results from revised simulations are shown. In both cases, the respective results for the narrow beam differ from those shown in Fig. 4 of [26].

The second implicit assumption involved in the correction for lack of CPE is that the photon mass attenuation and mass absorption coefficients taken from the NIST data bases [28], [34] are the same values as would be obtained from the interaction cross section data implemented in the MC codes. The associated uncertainty can be assumed to be mostly determined by the uncertainties of the interaction cross sections implemented in the codes which are in the order of a few percent [36].

The correction for lack of SPE, on the other hand, contains two potential sources for large uncertainties. One is related to the approximation of the fluence spectrum of all generations of scattered photons, assuming that the ratio of the total fluence to the fluence of the first generation is the same as that for the hypothetical case that photon scattering was isotropic. The second source of uncertainty is using only the information of the collision kerma rather than the full photon spectrum.

The uncertainty from the approximation used to estimate the fluence of higher-order scattered photons can be estimated by considering the change in the predicted collision kerma from first generation scattered photons when isotropic scattering is assumed. The results are shown in Table 1 as the ratio of the calculated collision kerma for isotropic scattering to that obtained by assuming angular distributions according to the Thomson formula and the Klein-Nishina formula for coherent (Rayleigh) and incoherent (Compton) scattering [32].

For the 50 kVp spectrum, this ratio is 44% for water and 45% for gold. This suggests that if the same factor would also apply to all generations of scattered photons, the fluence contribution from higher order scattered photons could be up to a factor of about 2.2 higher. Or, in other words, that the estimated contribution from higher generations of scattered photons may have an uncertainty as high as 55%, if one assumes the deviation of aforementioned ratio from unity to define the 95% confidence interval. Together with the fraction of the collision kerma coming from higher order scattered photons (listed in Table 2), the resulting overall relative uncertainty for the estimated collision kerma is in the order of 23% for the 50 kVp spectrum (Table 3). For the 100 kVp spectrum this overall uncertainty contribution is almost a factor of 2 lower. The values for the 60 kVp X-ray spectrum that are also shown in Table 1, Table 2 and Table 3 were derived from simulations performed for this spectrum by participant G1 in the same way as those reported in [26] for the other two radiation qualities.

To estimate the second uncertainty contribution, that comes from using only the condensed information of the collision kerma instead of the detailed spectrum, we consider the superposition of two photon energy fluence spectra where the resulting dose contribution from photon interactions with the GNP is known. For this purpose, we assume the hypothetical case that the fluence spectrum of all generations of scattered photons for both the 60 kVp and 100 kVp primary photon spectra is given by the 50 kVp X-ray spectrum multiplied by a factor. This scaling factor is chosen such that the extra collision kerma per fluence of primary photons has the same value as was obtained with the estimated spectrum from section 2.3.

**Table 1** Ratios of the collision kermas due to interactions of the first generation of scattered photons calculated for isotropic scattering and for the angular distributions from the Klein-Nishina and Thomson cross sections for incoherent and coherent photon scattering. The values were calculated from Table 3 in [32]. (The 50 kVp and 100 kVp spectra in [32] were the same as those used in [26].)

| spectrum | water | gold |
|---|---|---|
| 50 kVp | 0.44 | 0.45 |
| 60 kVp | 0.51 | 0.52 |
| 100 kVp | 0.91 | 0.81 |

**Table 2** Estimated relative contribution of higher order scattered photons to the total collision kerma, calculated from Table 3 in [32].

| spectrum | water | gold |
|---|---|---|
| 50 kVp | 0.41 | 0.42 |
| 60 kVp | 0.48 | 0.49 |
| 100 kVp | 0.64 | 0.65 |

**Table 3** Estimated relative uncertainty ($k = 2$) of the total collision kerma related to the sum spectrum of primary and all generations of scattered photons. These estimates are obtained by multiplying the deviation of the entries in Table 1 from unity with the respective entries in Table 2.

| spectrum | water | gold |
|---|---|---|
| 50 kVp | 0.23 | 0.23 |
| 60 kVp | 0.24 | 0.23 |
| 100 kVp | 0.06 | 0.13 |

The results obtained are shown in Fig. 5(a). The collision kerma in the GNP is symbolized by the downward triangles for the case of the primary 100 kVp photon spectrum. The corrected values according to section 2.3 and the assumed scaled 50 kVp photon spectrum are indicated by the open and full symbols, respectively. The ratio of the two quantities (shown as open symbols in Fig. 5(b)) is only slightly varying with distance from the GNP and deviates from unity by up to 25%. Using the same approach for the 60 kVp spectrum, and substituting a scaled 50 kVp spectrum for the scattered photon spectrum as obtained from Section 2.3, leads to a ratio of the collision kermas shown by the filled symbols in Fig. 5(b), which deviates from unity by only about 10%. For the collision kerma in water, the ratio of the values obtained with the estimated scattered photon spectrum is close to unity (dashed line Fig. 5(b)).

The deviations from unity of the data shown in Fig. 5(b) can be taken as estimate for the uncertainty due to a potential variation of the fluence spectrum of higher-order scattered photons with respect to the first generation of scattered photons. Using this approach, the overall combined uncertainty due to the correction for SPE is estimated to be about 25% for both of the considered X-ray spectra. For the 50 kVp primary photon spectrum, the dominant uncertainty contribution is given in Table 3, whereas for the 100 kVp spectrum the potential change in the photon spectrum dominates.

### 3.3. Uncertainty of the estimated DERs for an extended field.

Fig. 6 shows the results for the extended beam presented as deviation from unity of the DER for irradiation under secondary particle equilibrium. The symbols are the same data as in Fig. 4, while the dot-dashed line shows the mean of the DERs coming from data that passed the plausibility tests, which were marked by full symbols in Fig. 4. This mean constitutes the best estimate of the DER under secondary particle equilibrium for the four combinations of GNP size and photon spectrum.

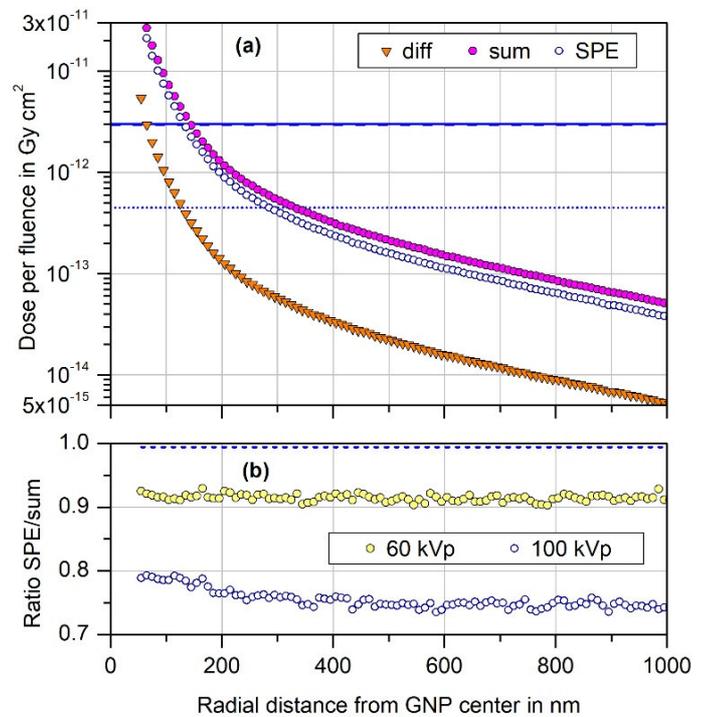

**Fig. 5:** (a) Difference between the simulation results of participant G1 for a 100 nm GNP present or absent in water traversed by a circular 100 kVp photon beam of diameter 110 nm (downward triangles). The open circles are the corrected values obtained by estimating the contribution of scattered photons with the procedure described in Section 2.3. The filled circles are obtained if the scattered photon spectrum is given by the primary 50 kVp spectrum multiplied by a factor such as to give the same collision kerma in gold as for the open circles. The horizontal lines indicate the dose per fluence to water if CPE (dotted) and SPE (solid, dashed) exist. (b) Radial variation of the ratio of the excess dose per fluence due to photon interactions with the GNP calculated with the estimated scattered photon spectrum to the one obtained with the scaled 50 kVp spectrum. Open circles indicate the ratio of the data shown in the upper part of the figure as open and filled circles. Filled circles indicate the corresponding ratio that is obtained for the 60 kVp primary spectrum when using a scaled 50 kVp spectrum instead of the scattered photon spectrum obtained with the procedure in Section 2.3. The dashed lines indicates the ratio of the values of dose per fluence to water obtained with the scattered photon spectrum according to Section 2.3 to that obtained with the scaled 50 kVp spectrum, both for the case of the 100 kVp primary spectrum.

All extended-beam data marked by full symbols in Fig. 4 agree within significantly less than 25% with their mean value so that the additional uncertainty coming from the scatter of results from different participants does not significantly increase the uncertainty of the DER (as long as those results that fail the plausibility tests are excluded). The grey shaded area represents the uncertainty band for 95% coverage with this slightly increased uncertainty.

The presentation of the deviation from unity of the DER in Fig. 6 is better suited for appreciating the range where significant dose enhancement occurs. It is perhaps a more appropriate way of presenting the results, as a GNP leads to additional energy deposition in its vicinity and the DER is by definition unity in absence of a GNP. Furthermore, it elucidates the fact that the datasets that do not pass the plausibility checks seem to have almost constant offsets from the dot-dashed line in the log-log presentation, i.e. the deviations from unity of the DERs have an almost constant ratio over the radial range shown. This supports the idea that the spread of all results would be significantly smaller if the normalization problems could be identified and corrected.

## 4. Discussion

### 4.1. Energy deposition from photon interaction in a GNP

Results presented in the previous Section demonstrate that simulations of the microscopic DER around a GNP by different Monte Carlo codes agree reasonably well when the simulation results pass the checks for proper implementation of the simulation setup. These consistency checks require quantities to be correctly normalized to primary particle fluence and to the event frequency of photon interactions (EFPI) in the GNP. Such checks are therefore based on the fundamental physical principle of energy conservation and can, in principle, also be used to predict absorbed dose enhancement from GNPs over larger scales such as cells or tissue voxels.

Normalization of the difference between simulation results with and without the presence of a GNP to the EFPI in a GNP also helps reveal fundamental aspects of the physical effects of GNPs in irradiated tissue. This is illustrated in Fig. 8, where the radial dependence of the average excess energy imparted in 10 nm-thick spherical shells around a GNP normalized to the EFPI is shown for the four combinations of GNP size and X-ray spectrum (averaged over results that passed the consistency checks). Therefore, data shown in Fig. 8 indicate the mean imparted energy around a GNP that experienced a single photon interaction.

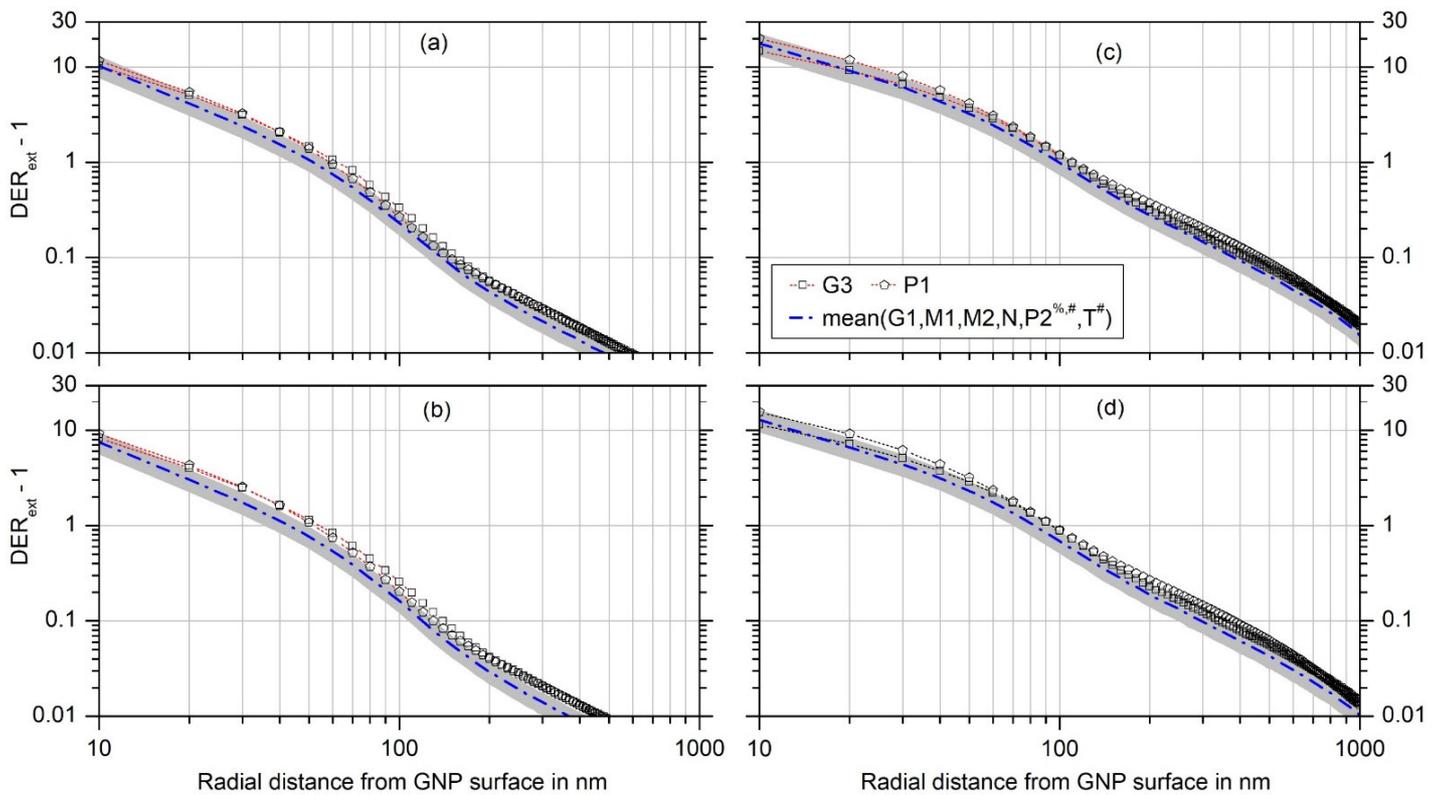

**Fig. 6**: Deviation from unity of the dose enhancement ratio (DER) for an extended photon field under secondary particle equilibrium for (a) a 50 nm GNP and 50 kVp X-rays, (b) a 50 nm GNP and 100 kVp X-rays, (c) a 100 nm GNP and 50 kVp X-rays and (d) a 100 nm GNP and 100 kVp X-rays. The different symbols correspond to the different participants in the exercise. Open symbols indicate that the reported simulation results did not pass the plausibility checks and no explanation could be reported by the participant. The other data sets passed the plausibility tests, where in some cases identified normalization issues have been corrected for (indicated by a hashtag). The dot-dashed line is the arithmetic mean of all data derived from simulations results that passed the plausibility tests, and the grey shaded area indicates the estimated 95% uncertainty band considering the scatter of these data and the uncertainties of the procedure used for correcting the lack of secondary photon equilibrium in the simulations.

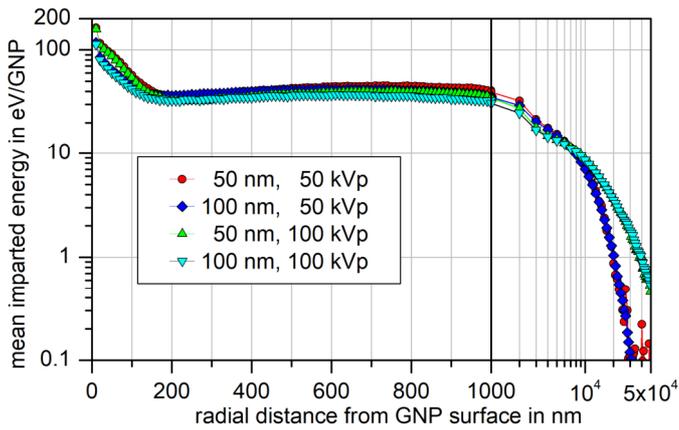

**Fig. 8:** Synopsis of the mean imparted energy in 10 nm-thick spherical shells around a GNP undergoing a photon interaction for two GNP sizes and photon energy spectra. The data points correspond to the means of participants' results that passed the consistency checks (i.e. data shown as dot-dashed lines in Fig. 6). The solid vertical line indicates a change from a linear to a logarithmic x-axis. In the right panel, the data points indicate the average over the 1 μm-thick spherical shells used in the simulations for radial distances exceeding 1000 nm.

Fig. 7 shows a close-up of the region of radial distances up to 1000 nm. Additional data included in this figure are the original results of participant P2, taking into account the photon spectra used in these simulations that are shown as grey and hatched areas in the inset of Fig. 7. These spectra correspond to the cumulative distributions of the 50 kVp and 100 kVp spectra used in the exercise (lines in the inset of Fig. 7).

In essence, the y-axis in Fig. 8 and Fig. 7 is the mean extra energy imparted in the spherical shells if a photon interaction occurred in the GNP. For large distances from the GNP, this mean energy imparted depends only on the X-ray spectrum and not on the GNP size (see right-hand panel of Fig. 8). This is to be expected as this energy deposition is from fluorescence photons or high-energetic electrons (produced by photo-absorption or Compton scattering in the GNP) that lose only a small fraction of their energy by interactions in the GNP prior to leaving it. (This small energy loss is in the order of the ratio of GNP size and electron CSDA range.)

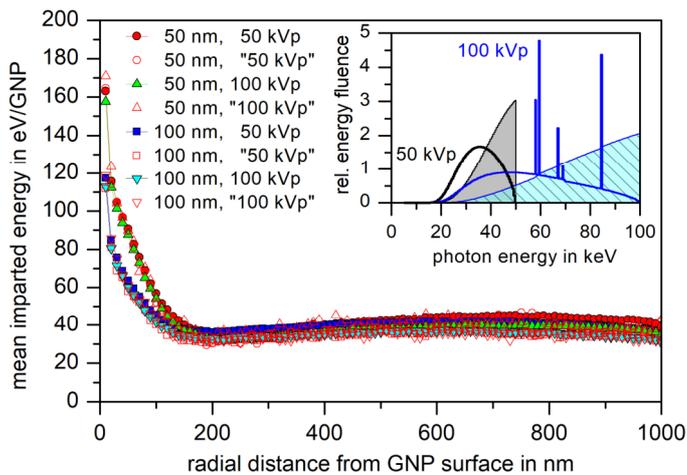

**Fig. 7:** Close-up of the mean imparted energy in 10 nm-thick spherical shells around a GNP that undergoes a photon interaction. In addition to the data shown in Fig. 8, the data derived from the original results of participant P2 are also included (open symbols). The inset shows the ratio of energy and particle fluence for the respective photon sources (shaded areas) compared to the spectra required in the exercise (lines).

For small radial distances up to about 100 nm, the situation is reversed such that the mean imparted energy depends on the GNP size and is almost independent of the photon spectrum. In this radial range, the energy is deposited by Auger electrons from M-shell vacancies (about 5 electrons on average [37]) or from L shell vacancies filled from another L shell (about 0.9 [37]). These electrons have energies up to about 3.5 keV, but more than 90% have energies below about 2.2 keV. Power-law extrapolation of values in the ESTAR database [38] gives ranges in the continuous slowing down approximation (CSDA) for electrons in gold of about 75 nm at 3.5 keV and about 35 nm at 2.2 keV.

The mean distance from a point inside a GNP to the surface is 60 nm and 30 nm for a 100 nm and 50 nm-diameter GNP, respectively. Therefore, in a 100 nm GNP, a larger fraction of these electrons is stopped before leaving the GNP than for a 50 nm GNP. This is confirmed by Table 4, that lists the mean numbers of electrons ejected per photon interaction in the GNP and the ratios of these values for the 100 nm and 50 nm GNPs. For a 100 nm GNP, the number of emitted electrons with energies below 2.2 keV is only 60% of those obtained for a 50 nm GNP. Hence, the comparatively smaller energy deposition around a 100 nm GNP compared to a 50 nm GNP, when normalized to the EFPI, is expected. Extrapolation of the ESTAR data for water gives CSDA ranges of 150 nm for 2.2 keV electrons and 44 nm for 1 keV. Therefore, the range of enhanced energy deposition is expected to be in the order of 100 nm as is observed in Fig. 8 and Fig. 7.

The other Auger electrons from L-shell vacancies in gold (about 1.3 on average [37]) have energies between 6 keV and 14 keV. The estimated CSDA ranges in gold are between 180 nm and 700 nm such that a fractional energy loss between 10% and 20% is expected for a 100 nm GNP and between 5% and 10% for a 50 nm GNP. The residual CSDA range in water of the electrons leaving the GNP is then expected to be between 1 μm and 5 μm. As the energy of most of these Auger electrons is in the range where the CSDA approximation is applicable, it is also expected that there is a range of radial distances where the excess energy imparted is constant with only a slightly higher value for the 50 nm GNP.

**Table 4** Mean number of electrons leaving the GNP normalized to the EFPI in the GNP and ratios of the values for 100 nm and 50 nm GNPs (last three rows). The values are given as the mean and the sample standard deviation of all simulations that passed the consistency checks.

|  | 50 kVp | | 100 kVp | |
| --- | --- | --- | --- | --- |
|  | 50 nm | 100 nm | 50 nm | 100 nm |
| all $T$ | 2.41 ± 0.48 | 1.96 ± 0.30 | 2.38 ± 0.54 | 1.92 ± 0.35 |
| $T \leq 2.2$ keV | 0.96 ± 0.45 | 0.57 ± 0.27 | 0.97 ± 0.51 | 0.58 ± 0.30 |
| $T > 2.2$ keV | 1.45 ± 0.05 | 1.39 ± 0.06 | 1.41 ± 0.04 | 1.34 ± 0.06 |
| all $T$ | 0.82 ± 0.03 | | 0.81 ± 0.03 | |
| $T \leq 2.2$ keV | 0.60 ± 0.03 | | 0.60 ± 0.03 | |
| $T > 2.2$ keV | 0.96 ± 0.01 | | 0.95 ± 0.03 | |

*4.2. Estimated DER from GNPs in larger volumes*

Fig. 8 and Fig. 7 show that the energy deposition in the first 100 nm around a GNP experiencing a photon interaction from a low-energetic X-ray spectrum depends mainly on the GNP size. This energy deposition is almost independent of both GNP size and photon spectrum in the range between 100 nm and 1 μm. Therefore, the average local dosimetric effect of a GNP is given by the imparted energy due to a photon interaction multiplied by the EFPI in a GNP. The EFPI depends on the photon fluence at the location of the GNP. For a single GNP, it is proportional to the absorbed dose to water in its absence.

For a GNP in a volume of tissue loaded with nanoparticles, the photon fluence is enhanced by the contribution of photons scattered by GNPs. The additional fluence due to these scattered photons depends on the concentration of gold in the volume loaded with GNPs as well as on the size of this volume (which may be an explanation for the apparent increase of DER with field size concluded by Vlastou at al. [21].) Thus, a general prediction of the EFPI is not possible, as it depends on the size of the volume containing GNPs as well as their concentration (i.e. number per volume).

The radial integral $\Delta E^*_{d,g}$ of the imparted energy around a GNP per photon interaction (shown in Fig. 8 and Fig. 7) can be used to estimate the DER for the absorbed dose averaged over such larger regions as follows:

$$DER = 1 + x_g \frac{\bar{n}_g(\Phi)}{\rho_{Au} V_{np}} \frac{\Delta E^*_{d,g}}{D_w} \left(1 - f_{np}(x_g)\right) \quad (8)$$

Here, $x_g$ is the MFG, $\bar{n}_g$ is the fluence-dependent EFPI, $\rho_{Au}$ is the mass density of gold, $V_{np}$ is the GNP volume and $D_w$ is the absorbed dose to water in the absence of GNPs. $f_{np}$ is the fraction of the energy transported out of the GNP that is absorbed by interactions of the respective electrons within other GNPs. The term $\bar{n}_g/(\rho_{Au} V_{np})$ is the photon-spectrum average of the mass attenuation coefficient of gold times the photon fluence and, therefore, independent on the GNP size. Therefore, for small values of $x_g$, the deviation from unity of the DER is proportional to $x_g$, as the fractional energy loss to other GNPs is expected to be negligible.

With increasing MFG, the energy absorbed in other GNPs increases. The resulting decrease of the deviation of the DER from unity is, however, counteracted by an increase of the EFPI in a GNP due to the additional fluence of photons scattered at GNPs. The net effect is difficult to predict in general. However, an estimate can be obtained from eq. (8) by a) using the photon fluence in the absence of GNPs to calculate the interaction frequency and b) neglecting the term $f_{np}$. The result can be expected to provide a reasonable approximation of the average DER for a concentration of GNPs in an irradiated volume.

*4.3. Comparison with DERs reported in literature*

The consistency checks mentioned in Section 2.4 are, in essence, based on the fundamental principle of energy conservation and thus enable a critical assessment of results found in literature. This was conducted for a selection of papers whose studies overlapped with our work (EURADOS exercise) with respect to the size of GNPs and the spectral range covered by the photon sources used in the simulations.

Generally, reported DER values for photon spectra and GNP sizes similar to those used in the EURADOS exercise vary from negligible deviations from unity when averaged over cell compartments [16] to several hundred over micrometer distances from a single nanoparticle [12]. Leung et al. even reported values of order 2000 for 50 kVp X-rays and GNP sizes up to 100 nm for what they called "interaction enhancement ratio" [13] .

Jones et al. [12] determined what they called microscopic DER (mDER) by taking the ratio of dose point kernels, i.e. the energy deposited per mass around a nanoparticle of gold or of water by the electrons produced by photon interactions inside the respective nanoparticle. The same approach was also used in a study by Lin et al. [14] to determine DERs for X-rays and protons. Thus, the simulation results in both papers are not only lacking lateral but also longitudinal secondary particle equilibrium. As the volume contributing to the energy imparted in the absence of the GNP was by a factor of 1000 (100 nm GNP) or 2000 (50 nm GNP) smaller than in our simulations, the deviation of the DERs for our scoring geometry would be enhanced by the same factors. However, Jones at al. [12] used larger spherical shells which leads to a reduction of the dose ratios. Therefore, their mDERs are consistent with the DERs from our simulation results, although the magnitude of both values is not the expected DER around GNPs in a realistic irradiation.

Jones at al. [12] also used their dose point kernels to assess the dose around GNPs from transmission electron microscope (TEM) images of GNP distributions in cells. At first glance, their reported DERs of about 1.5 at about 10 μm distance from GNP clusters for 50 kVp irradiation seem to contradict our results. However, Jones at al. [12] only performed a two-dimensional integral. Their explicit assumption "that the GNP distribution in three dimensions is similar to that of the two-dimensional image" implies that they were effectively calculating the DER of gold nanorods of infinite length and not for GNPs. Thus, their microscopic DER is overestimated due to the enhanced MFG that is implicitly assumed.

Leung et al. [13] used a similar irradiation geometry as in the exercise (photon beam cross section equal to GNP cross section) and studied relative dose variation around the GNPs for six different beam qualities and GNP sizes from 2 nm to 100 nm. Their Fig. 4a shows results normalized to the highest value for the 100 nm GNP, and hence, a quantitative comparison to our results is not possible. Their relative dose appears to scale approximately with the GNP diameter, which is to be expected from eq. (7) (considering that for the simulation geometry the fluence is proportional to $1/r_g^2$). Their "interaction enhancement ratio" shown in Fig. 6 of [13] is the ratio of interaction events producing secondary electrons in a gold and a water nanoparticle, respectively, and has values of about 2000 for the 50 kVp spectrum and all GNP sizes. This quantity is essentially the product of the ratios of a) the densities (about 20) and b) the mass absorption coefficients (about two orders of magnitude) of gold compared to water in this photon energy range. The number of interaction events shown in their Fig. 5 has approximately the expected linear dependence on GNP diameter (distorted to a logarithm curve due to the linear-log presentation in their figure).

In the frame of a comparison of the influence of different cross-section datasets in Geant4, Sakata et al. [39] studied dose enhancement produced by a parallel beam of 10 keV electrons impinging on a GNP in water compared to a water nanoparticle and found DERs in the order of 100 at distances between 10 µm and 1 cm from the GNP. As 10 keV electrons have a mean range in water of about 2.5 µm [38], this study does not relate to any realistic irradiation scenario, and the simulation setup lacks secondary particle equilibrium.

Gadoue and Toomeh [40] used a deterministic code to study the enhancement of linear energy transfer and absorbed dose around single GNPs and small clusters of GNPs of 50 nm and 100 nm diameter. Their DERs for a 120 kVp X-ray spectrum in the first 25 nm around the GNPs were between about 6 and 12 for the 50 nm GNP and about 10 and 17 for the 100 nm GNP. These values are similar to those shown in Fig. 4 of this paper for the 100 kVp spectrum, and thus support the approach and results of this work.

Douglass et al. [16], [41] determined the enhancement of the average dose in different cellular compartments induced by GNPs. In their simulation, the region of interest (a cube of 400 µm side) contained 850 cells of idealized geometry, and the photon field was 1x1 mm² so that lateral secondary charged particle equilibrium can be assumed for a 80 kVp X-ray irradiation [32]. Two cases were considered: a) a cluster of GNPs in the cytoplasm that was modeled by a 400 nm-diameter gold sphere; b) GNPs aggregating at the cell nucleus membrane, modelled as a 300 nm-thick gold layer covering the nuclear membrane. In the first case, no significant dose enhancement was found. For the second case, however, the DER averaged over the cell was about 15 while the DER for the cell nucleus was about 56.

It has to be noted that in the cell model of Douglass et al. [41], the nucleus has a surface of more than 100 µm² such that the mass of gold in the second case is about three orders of magnitude higher than in the first geometry (a cluster of GNPs in the cytoplasm). In fact, the mass fraction for the case of GNPs "aggregated" at the nucleus membrane amounts to between 15% and 30% (depending on the cell dimensions). This appears to be unrealistic as the highest reported MFG in tumor cells is about 1.8% [42]. (Besides, a cell with a 300 nm gold enclosure around its nucleus would be deprived of its vital functions.)

For an idealized clinical tumor irradiation, Koger and Kirkby [18] studied the dependence of the DER on irradiation geometries, MFG and photon energy. Their study encompassed a variety of monoenergetic photon beams as well as several X-ray spectra (including 100 kVp) and linac radiation. GNP sizes of 2 nm and 20 nm GNPs were studied applying a procedure for correcting dose values obtained by simulations in a mixture of gold and water to the absorbed dose to medium around GNPs [43]. Results were reported as macroscopic DERs defined as dose to tumor tissue with and without GNPs present [15].

For X-ray spectra, these doses are approximately equal to the average dose within a cell. For higher energy photons, however, this equivalence only holds if the photon beam is large enough to assure secondary charged particle equilibrium.

The average DER obtained from eq. (8), by neglecting the term $f_{np}$ and calculating the EFPI with the photon fluence in water for all generations of scattered photons in a depth of 100 µm for the 100 kVp spectrum, is about 3. Almost the same value is obtained for a depth in water of 2.5 cm, which is closer to the geometrical setup of the simulations by Koger and Kirkby [18]. Their DER results for a 100 kVp spectrum are between 1.8 and 2.6. The discrepancy can be attributed to the fact that our estimate for all generations of scattered photons is for a laterally infinite photon field, while their results were obtained from simulations using a finite photon field.

The approach of Koger and Kirkby [43] avoids the problem that straightforward simulations of a macroscopic irradiation cannot achieve primary particle fluences that assure sufficient interactions with GNPs. For example, with the 50 kVp spectrum, a photon fluence of $10^8$ cm$^{-2}$ corresponds to an EFPI in a 100 nm GNP of $3\times10^{-5}$ (and to an absorbed dose to water of $7\times10^{-5}$ Gy). In a recent review paper, Vlastou et al. speculated that this computational challenge may be the reason for the lack of consistent evidence regarding GNP effects [21].

Nevertheless, three studies have tried such straightforward simulation using the MCNP code and a hierarchical geometry with cubic voxels of decreasing size [17], [20], [44]. Spherical GNPs were placed in the centers of the lowest-level voxels that were filled with water. Energy deposition in the water volumes was scored in higher-level voxels of 1 mm³ or 8 mm³ size.

All three papers report an increase in the average DER with increasing GNP size for a constant MFG [17], [20], [44]. This dependence on GNP size appears to be an artefact of the simulation, as the results are not compatible with what one would expect from energy conservation: As was shown by Koger and Kirkby [43], the fraction of the energy transferred to electrons by soft X-ray photon interactions inside a GNP and imparted to water by these electrons decreases with increasing GNP size. This dependence is expected, as was discussed in section 4.1, and so is a decrease of the DER with increasing GNP size. Furthermore, most of DERs for 100 nm GNPs reported in [17] are close to or even exceeding the DER expected for a homogenous mixture using the mass energy absorption coefficients [28]. The same applies to all DERs reported in [20], including those for smaller GNP dimensions.

Even though the three papers show a graphical illustration of the geometry construction, essential information is missing as to how the MFG was actually realized in the simulation setup. Mesbahi et al. [17] studied MFG of 7 mg/g and 18 mg/g and used, for example, 100 nm GNPs in 1 µm³ voxels which corresponds to a mass fraction of 10 mg/g. While the lower mass fraction (which was also used in [20]) could be obtained by leaving 30% of the 1 µm³ voxels without GNP, it is not described how the 18 mg/g was achieved. Kheshevarz et al. [44] refer to the work of Mesbahi et al. [17], but did not include the information about the smallest voxel size. In their work, the difference between uniform and non-uniform distribution of GNPs was studied, where the DERs for soft X-rays and the uniform distribution of 100 nm GNPs are compatible with what is expected from the mass-energy absorption coefficients and the results of Koger and Kirkby [43].

Kakade and Sharma [19] studied dose enhancement from 7 mg/g and 18 mg/g gold in tumor tissue and in a gel dosimeter. As a GNP size is not specified, one can assume that they used a homogeneous mixture of gold with the gel and tissue material, respectively. Their DERs are in agreement with the prediction based on mass-energy transfer coefficients from [28].

In a recent study by Mirza et al. [45], dose enhancement in radiochromic films coated with gold nanofilms of 20 to 100 nm thickness was determined experimentally. Depending on the film thickness, the DERs obtained for 50 kVp irradiation ranged from 2.1 to 6.1 when the dose was assessed by optical readout, and from 2.6 to 4.6 when Raman spectroscopy was used. Detailed analysis was performed for the comparison of their results with expected values based on the physical processes involved in the dose deposition (similar to our discussion in sections 4.1 and 4.2). If the whole active layer thickness of the film was used, the predicted DERs were smaller than the measured ones, while they were much higher than measured DERs when the thickness of the detection volume was defined by the ranges of secondary electrons [45].

Beyond the in-depth discussion given by Mirza et al., it is worth noting that their observed increase of DER with gold film thickness is a consequence of the increasing photon interaction frequency within the films. Furthermore, simplifications in their approach, such as using ranges for electrons of the average electron energy, may have contributed to the discrepancies between their experiment and theoretical results. In addition, the choice of volume for scoring the dose together with the film thickness essentially defines a 'mass fraction' of gold. Using the respective value obtained for the 50 nm film and the 25 µm active layer thickness and taking the other input values for eq. (8) from our data for the 50 nm GNP and 50 kVp X-rays gives an estimated DER about 3.5 which is in agreement with the experimental values reported in [45].

*4.4. Limitations of the concept of a local DER*

In principle, the rationale behind eq. (8) can also be used for estimating local dose enhancement at different locations inside or around a volume loaded with GNPs. Using an approach similar to ours, Lin et al. estimated the DER in the walls of blood vessels with nanometric voxel resolution [46]. In this case, the simulation result is not the absorbed dose but rather the microdosimetric specific energy, i.e. a stochastic quantity. This stochasticity resulted in what Lin et al. called "dose spikes" that were particularly prominent when GNPs were always attached to the vessel walls. The explanation for such "dose spikes" is that there are photon interactions in GNPs in close proximity to the affected voxels and that the electrons are emitted from the GNP in the direction of these voxels.

This stochasticity implies that enhancement of the absorbed dose may not be the appropriate quantity for describing the local radiation effects of GNPs [32], [47]. To illustrate this point, Fig. 9 shows a comparison of the expected total mean imparted energy in 10 nm-thick spherical shells around a 50 nm GNP (small symbols connected by a solid line) compared with the respective values of mean imparted energy for zero (blue line), one (large filled circles) or two (open diamonds) photon interactions in the GNP. This data, together with the EFPI values given in the legend, correspond to an absorbed dose of 2 Gy delivered by the 50 kVp primary photon spectrum (and a low concentration of GNPs so that the additional fluence of photons scattered on GNPs is negligible). Hence, the blue line is also the mean imparted energy in the absence of a GNP that increases quadratically with increasing radial distance.

The important observation in Fig. 9 is that there is a mean imparted energy of several hundreds of eV in the first 10 nm-thick shell around the GNP in the cases of one, two or more photon interactions in the GNP. Assuming a mean energy per ionization of 25 eV, this corresponds to a number of ionizations in the order of 10 in close vicinity of the GNP. This suggests a possible explanation for the radio-sensitizing effects of GNPs (much larger increase of biological effectiveness compared to the DER [22]): These localized ionizations may destroy the coating of the GNP that is essential for its biocompatibility.

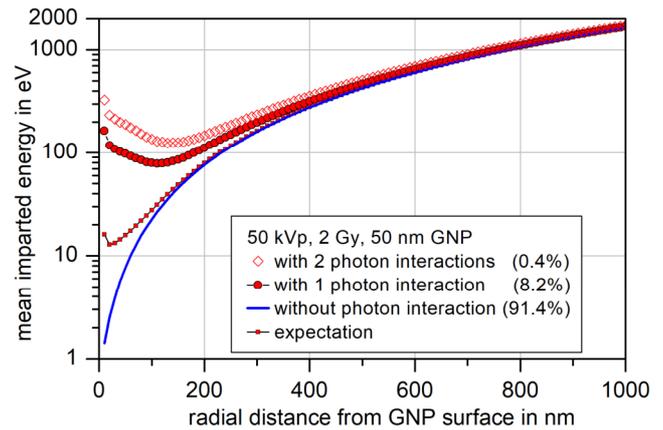

**Fig. 9:** Expected mean imparted energies in 10 nm-thick radial shells around a 50 nm GNP irradiated by the 50 kVp spectrum at a particle fluence producing an absorbed dose of 2 Gy in the absence of the GNP. The small full symbols indicate the average imparted energy around the GNP, and the solid line indicates the mean imparted energy in the absence of the GNP. The data for a GNP experiencing one or two photon interactions are shown as filled and open symbols respectively.

Furthermore, if the GNP experiences a photon interaction and is located, for example, at the cell nucleus membrane (or the walls of a blood vessel), a local destruction of the nucleus membrane may occur. The number of electrons emitted from a GNP and the average energy imparted by these electrons are proportional to the number of photon interactions in the GNP [46]. Thus, GNPs with more than one photon interaction have a higher potential to induce such damage. The important point here is that the frequency of occurrence of a number of photon interactions in a GNP increases with photon fluence according to a Poisson distribution. Thus, the probabilities for more than one photon interaction increase with the respective power of fluence, i.e. super-linearly.

Therefore, there may be a physical contribution to the radio-sensitization that occurs in addition to an enhancement of the absorbed dose by GNPs [48]. This contribution may be due to the (small) fraction of GNPs that undergo more than one photon interaction which has a non-linear dose dependence. It seems that this has not been considered in the literature so far. The DER, on the other hand, is invariant with absorbed dose and, thus, not reflecting aforementioned effects.

In addition, it is worth noting that the number density of GNPs is inversely proportional to the GNP volume, while the EFPI in a GNP is directly proportional to the volume. Thus, for a given MFG, the event frequency of photon interactions with gold per volume of tissue is constant [49]. Accordingly, the number density of GNPs with exactly one photon interaction increases with decreasing size of the GNP. On the contrary, the number densities of GNPs interacting with *n* photons increase with GNP size while EFPI is below *n*-1 and decrease with GNP size for higher event frequencies. This implies that any biological effects triggered by the local energy deposition close to the surface of a GNP after photon interaction have a complex, yet physically understandable dependence on GNP size (and on MFG, photon fluence and photon spectrum). It should be noted that such dependence cannot be easily assessed by deterministic approaches as the one presented by Gadoue et al. [40].

# 5. Conclusion

In [26] an intermediate state of the analysis of the code intercomparison exercise was shown with a focus on elucidating potential differences between codes. For this reason, DER results were shown that were directly calculated from the simulation results. In this work, we have used an approximate correction procedure to derive the DERs that would be found for an extended photon field where secondary particle equilibrium is ensured. These DERs are significantly deviating from unity only within the first few hundred nanometers from the GNP surface and have maximum values in the order of magnitude 10 (Fig. 4).

In addition, an estimate of the uncertainty related to the corrections applied has been presented that amounts to about 25% of the deviation of the DER from unity (at 95% confidence level). After exclusion of datasets with remaining normalization issues, the scatter between different participants' results does not significantly increase this uncertainty. However, this uncertainty is about an order of magnitude higher than what is generally considered acceptable for dosimetry in radiotherapy. A follow-up code comparison exercise is therefore in preparation to fully assess the uncertainty based on deviations of results between codes for simulations ensuring secondary particle equilibrium.

The need for such an extended intercomparison is supported by the findings in our critical discussion of literature. As discussed above, several studies used approaches that failed to ensure secondary particle equilibrium and, consequently, the giant DERs reported in these studies can be attributed to this deficiency in the simulation setups. Large DER values from average absorbed dose over cells or larger tissue volumes appear to be the result of a high mass fraction of gold. In some cases, the reported dependence of DER on the GNP size was contrary to what would be expected from energy conservation. Such results should therefore be treated with caution.

Finally, the importance of stochasticity for radiation effects of GNPs has not yet been thoroughly investigated [47]. It may provide an explanation for apparent contradictions between the occurrence of large biological radiation effects of GNPs in situations where simulations predict much smaller DERs (i.e. for average absorbed dose). As suggested by Moradi et al. [22] Monte Carlo simulations have carved out this contradiction in DER results and, thus, highlighted the importance of chemical or biological effects of GNPs. While such effects are important, the role of physical effects is not yet fully understood. The challenge and future role of Monte Carlo simulations of GNP radiation effects is to determine the impact of stochastic effects with respect to different GNP sizes and radiation spectra.

# Acknowledgements

This work was, in part, funded by DFG (grant nos. 336532926 and 386872118) and the National Cancer Institute (grant no. R01 CA187003). Werner Friedland is acknowledged for providing his simulation results without claiming co-authorship.